\documentclass[aps,pra,reprint,superscriptaddress,nofootinbib]{revtex4-2}

\usepackage{amsmath,amssymb,amsfonts,bm}
\usepackage{graphicx}
\usepackage{braket}
\usepackage{hyperref}

\begin{document}

\title{Measurement-Conditioned Bloch-Sphere State Steering with a Continuous-Variable Pointer}

\author{Necati Celik }
\affiliation{Department of Physics Engineering, Faculty of Engineering and Natural Sciences, G\"um\"u\c{s}hane University, G\"um\"u\c{s}hane, T\"urkiye}

\begin{abstract}
We investigate measurement-conditioned state steering of a qubit coupled to a continuous-variable pointer through a state-dependent displacement interaction. A position measurement on the pointer induces a nonunitary conditional transformation of the qubit, allowing its post-measurement state to move along a trajectory on the Bloch sphere. We derive the conditional state, the measurement-outcome distribution, and the corresponding Bloch-sphere coordinates. For a real Gaussian pointer, the conditional state follows a meridian, whereas a phase-engineered Gaussian pointer produces a trajectory with a measurement-dependent azimuthal angle. We further describe outcome-dependent unitary feedback for preparing a prescribed target state from a known initial qubit state. We distinguish this state-preparation task from implementing a universal deterministic single-qubit gate, which does not follow from the proposed interaction and measurement alone. Finally, we derive the quantum Fisher information of the joint qubit--pointer state for estimating the displacement-generating parameter and discuss the implications of pointer squeezing, phase engineering, and experimental implementation. The results provide a framework for analysing conditional Bloch-sphere steering using continuous-variable measurement resources.
\end{abstract}

\maketitle

\section{Introduction}

Quantum measurements can modify the state of a quantum system as well as provide information about it. When a discrete-variable system interacts with a continuous-variable degree of freedom, a measurement of the latter can induce a state-dependent transformation of the former. Such interactions are central to quantum measurement theory, quantum control, and measurement-based protocols \cite{wiseman2009quantum,proctor2017ancilla,jordan2006qubit}.

A useful example is a qubit coupled to a pointer through a conditional displacement. The qubit states become correlated with displaced pointer wavefunctions, and a subsequent position measurement generates a conditional qubit state. Because the relative amplitudes and phases of the two qubit components depend on the measurement outcome, the conditional state can move along a trajectory on the Bloch sphere. This mechanism underlies measurement-induced state steering and is closely related to quantum trajectory dynamics under measurement backaction \cite{murch2013observing,roy2020measurement}.

The purpose of this work is to formulate this process as a measurement-conditioned state-steering protocol. We derive the conditional Bloch-sphere coordinates for general pointer wavefunctions and examine Gaussian and phase-engineered Gaussian pointers. Gaussian pointer states are a natural continuous-variable resource \cite{weedbrook2012gaussian}. We also consider outcome-dependent feedback for preparing a target state from a known initial state; feedback based on conditional unitary operations after a measurement is well established in quantum control \cite{wiseman2009quantum,jordan2006qubit}. Such conditional displacements are also relevant to trapped-ion and circuit-QED platforms, where state-dependent forces on motional or bosonic modes have been demonstrated \cite{leibfried2003quantum,meekhof1996generation}.

An important distinction is required between conditional state steering and universal quantum-gate implementation. The conditional operation generated by a pointer-position measurement is generally nonunitary. Consequently, an outcome-dependent unitary correction can map a known conditional pure state to a desired target state, but this does not establish that the same protocol implements an arbitrary unitary gate on an unknown input qubit. This distinction connects to broader analyses of nonunitary operations and feedforward in measurement-based quantum computation \cite{proctor2017ancilla,nielsen2006cluster}. We therefore focus on state preparation and steering rather than claiming a deterministic universal single-qubit gate.

We also analyse parameter estimation within the same interaction model. In particular, we calculate the quantum Fisher information (QFI) of the joint qubit--pointer state and clarify how the pointer's position and momentum uncertainties enter the estimation problem. The QFI provides the ultimate precision bound for parameter estimation \cite{paris2009quantum,pezze2018quantum,giovannetti2011advances}.
\section{Qubit--pointer interaction}

\subsection{System and interaction Hamiltonian}

Consider a qubit with ground state $\ket{g}$ and excited state $\ket{e}$. Its Hamiltonian is
\begin{equation}
\hat{H}_S=\epsilon\ket{e}\bra{e},
\label{eq:system_hamiltonian}
\end{equation}
where the ground-state energy is set to zero and $\epsilon$ is the energy separation.

The pointer is described by canonical position and momentum operators satisfying
\begin{equation}
[\hat{X},\hat{P}]=i\hbar.
\label{eq:canonical_commutator}
\end{equation}
We consider the interaction unitary
\begin{equation}
\hat{U}_{\epsilon}
=
\exp\left(
-\frac{i\lambda\epsilon}{\hbar}
\ket{e}\bra{e}\otimes\hat{P}
\right),
\label{eq:interaction_unitary}
\end{equation}
where $\lambda$ is a coupling-duration parameter. The product
\begin{equation}
d=\lambda\epsilon
\label{eq:displacement}
\end{equation}
has dimensions of length when $\hat{P}$ is the physical pointer momentum.

The operator $\ket{e}\bra{e}$ makes the displacement conditional on the qubit state. The ground-state component leaves the pointer unchanged, whereas the excited-state component displaces it by $d$.

We assume an initial product state
\begin{equation}
\ket{\Psi_0}
=
\ket{\psi_S}\otimes\ket{\psi_P},
\label{eq:initial_product_state}
\end{equation}
where
\begin{equation}
\ket{\psi_S}
=
\cos\frac{\theta_0}{2}\ket{g}
+
e^{i\phi_0}\sin\frac{\theta_0}{2}\ket{e}.
\label{eq:initial_qubit}
\end{equation}
The initial pointer wavefunction in the position representation is
\begin{equation}
\psi_P(x)=\braket{x|\psi_P},
\qquad
\int_{-\infty}^{\infty}|\psi_P(x)|^2\,dx=1.
\label{eq:pointer_normalization}
\end{equation}

\subsection{Joint state after the interaction}

Using the translation identity
\begin{equation}
\exp\left(-\frac{id\hat{P}}{\hbar}\right)\psi_P(x)
=
\psi_P(x-d),
\label{eq:translation_identity}
\end{equation}
the joint state after the interaction becomes
\begin{equation}
\begin{aligned}
\ket{\Psi}
={}&
\cos\frac{\theta_0}{2}\ket{g}\ket{\psi_P}\\
&+
e^{i\phi_0}\sin\frac{\theta_0}{2}
\ket{e}\ket{\psi_{P,d}},
\end{aligned}
\label{eq:joint_state}
\end{equation}
where
\begin{equation}
\braket{x|\psi_{P,d}}=\psi_P(x-d).
\label{eq:displaced_pointer}
\end{equation}
Equation~\eqref{eq:joint_state} shows that the interaction correlates the qubit basis states with two pointer states separated by the displacement $d$.

\section{Conditional qubit state and Bloch-sphere coordinates}

\subsection{Position measurement}

Suppose that the pointer position is measured and the outcome is $x$. The corresponding measurement operator acting on the qubit is
\begin{equation}
\hat{M}_x
=
\psi_P(x)\ket{g}\bra{g}
+
\psi_P(x-d)\ket{e}\bra{e}.
\label{eq:measurement_operator}
\end{equation}
In the ordered basis $\{\ket{g},\ket{e}\}$, this operator has the matrix representation
\begin{equation}
M_x=
\begin{pmatrix}
\psi_P(x) & 0\\
0 & \psi_P(x-d)
\end{pmatrix}.
\label{eq:measurement_matrix}
\end{equation}
It satisfies the completeness relation
\begin{equation}
\int_{-\infty}^{\infty}\hat{M}_x^\dagger\hat{M}_x\,dx
=
\hat{I}_S.
\label{eq:measurement_completeness}
\end{equation}

The unnormalized conditional qubit state is
\begin{equation}
\begin{aligned}
\ket{\widetilde{\psi}_{S|x}}
={}&
\cos\frac{\theta_0}{2}\psi_P(x)\ket{g}\\
&+
e^{i\phi_0}\sin\frac{\theta_0}{2}
\psi_P(x-d)\ket{e}.
\end{aligned}
\label{eq:unnormalized_conditional_state}
\end{equation}
Its squared norm gives the probability density of the measurement outcome:
\begin{equation}
\begin{aligned}
p(x)
={}&
\cos^2\frac{\theta_0}{2}|\psi_P(x)|^2\\
&+
\sin^2\frac{\theta_0}{2}|\psi_P(x-d)|^2.
\end{aligned}
\label{eq:outcome_probability}
\end{equation}
The normalized conditional state is therefore
\begin{equation}
\ket{\psi_{S|x}}
=
\frac{\ket{\widetilde{\psi}_{S|x}}}{\sqrt{p(x)}},
\label{eq:normalized_conditional_state}
\end{equation}
whenever $p(x)>0$.

The conditional transformation is generally nonunitary because the two diagonal entries of $\hat{M}_x$ need not have equal magnitudes. This nonunitarity is the mechanism responsible for the outcome-dependent change in the qubit amplitudes.

\subsection{Conditional polar and azimuthal angles}

We express the normalized conditional state in the standard Bloch-sphere form
\begin{equation}
\ket{\psi_{S|x}}
=
\cos\frac{\theta(x)}{2}\ket{g}
+
e^{i\phi(x)}
\sin\frac{\theta(x)}{2}\ket{e}.
\label{eq:conditional_bloch_state}
\end{equation}
Comparing Eqs.~\eqref{eq:unnormalized_conditional_state} and
\eqref{eq:conditional_bloch_state} gives
\begin{equation}
\tan\frac{\theta(x)}{2}
=
\tan\frac{\theta_0}{2}
\frac{|\psi_P(x-d)|}{|\psi_P(x)|},
\label{eq:conditional_polar_angle}
\end{equation}
and
\begin{equation}
\phi(x)
=
\phi_0+
\arg\psi_P(x-d)-\arg\psi_P(x)
\pmod{2\pi}.
\label{eq:conditional_azimuthal_angle}
\end{equation}
These expressions apply wherever the relevant amplitudes are nonzero; points at which one component vanishes can be treated by taking the corresponding limit.

The conditional Bloch vector is
\begin{equation}
\bm{r}(x)
=
\begin{pmatrix}
\sin\theta(x)\cos\phi(x)\\
\sin\theta(x)\sin\phi(x)\\
\cos\theta(x)
\end{pmatrix}.
\label{eq:conditional_bloch_vector}
\end{equation}
Equations~\eqref{eq:conditional_polar_angle} and
\eqref{eq:conditional_azimuthal_angle} show that the pointer's amplitude profile determines the conditional polar angle, whereas its spatial phase profile contributes to the conditional azimuthal angle.

The outcome $x$ is stochastic. Consequently, the interaction does not independently set the two Bloch angles to arbitrary prescribed values. Instead, it generates a probability distribution over conditional states along a trajectory determined by the pointer wavefunction.

\section{Gaussian pointer and meridional steering}

\subsection{Real Gaussian pointer}

Consider a normalized Gaussian pointer wavefunction,
\begin{equation}
\psi_P(x)
=
\frac{1}{(2\pi\sigma^2)^{1/4}}
\exp\left(-\frac{x^2}{4\sigma^2}\right),
\label{eq:gaussian_pointer}
\end{equation}
where $\sigma^2$ is the position variance. The pointer has zero mean position and momentum, with
\begin{equation}
(\Delta X)^2=\sigma^2,
\qquad
(\Delta P)^2=\frac{\hbar^2}{4\sigma^2}.
\label{eq:gaussian_uncertainties}
\end{equation}

For this real wavefunction, the phase of the pointer is independent of position. Equation~\eqref{eq:conditional_azimuthal_angle} therefore gives
\begin{equation}
\phi(x)=\phi_0.
\label{eq:gaussian_azimuthal_angle}
\end{equation}
The ratio of the displaced and undisplaced wavefunctions is
\begin{equation}
\frac{\psi_P(x-d)}{\psi_P(x)}
=
\exp\left[
\frac{d}{2\sigma^2}\left(x-\frac{d}{2}\right)
\right].
\label{eq:gaussian_amplitude_ratio}
\end{equation}
The conditional polar angle becomes
\begin{equation}
\tan\frac{\theta(x)}{2}
=
\tan\frac{\theta_0}{2}
\exp\left[
\frac{d}{2\sigma^2}\left(x-\frac{d}{2}\right)
\right].
\label{eq:gaussian_polar_angle}
\end{equation}

For $d\neq 0$ and an initial state with both basis-state amplitudes nonzero, the conditional polar angle approaches opposite poles as $x$ traverses the real line, with the direction of traversal determined by the sign of $d$. At finite $x$, the endpoints are approached rather than attained exactly. The conditional states thus lie on the meridian $\phi=\phi_0$.

The measurement-outcome distribution for this pointer is
\begin{equation}
\begin{aligned}
p(x)
={}&
\cos^2\frac{\theta_0}{2}
\frac{e^{-x^2/(2\sigma^2)}}{\sqrt{2\pi}\sigma}\\
&+
\sin^2\frac{\theta_0}{2}
\frac{e^{-(x-d)^2/(2\sigma^2)}}{\sqrt{2\pi}\sigma}.
\end{aligned}
\label{eq:gaussian_outcome_distribution}
\end{equation}
This distribution determines how frequently different conditional states are obtained. Although the formal trajectory can span nearly the full meridian, its extreme regions may have very small probabilities.

\subsection{Role of pointer width}

The parameter $\sigma$ controls the overlap of the two pointer wavefunctions. A large $\sigma$ produces substantial overlap and weak discrimination between the qubit basis states. A smaller $\sigma$ makes the displaced wavefunctions more distinguishable and increases the variation of the conditional amplitude ratio with $x$.

For a minimum-uncertainty Gaussian pointer, reducing $\sigma$ increases the momentum variance according to Eq.~\eqref{eq:gaussian_uncertainties}. Thus, enhanced sensitivity to the displacement is associated with a greater momentum-space resource. Comparisons between different pointer widths should account for this change in the available momentum variance.

\section{Phase-engineered pointer and nonmeridional trajectories}

\subsection{Chirped Gaussian pointer}

To introduce position-dependent phase differences, consider the normalized pointer wavefunction
\begin{equation}
\psi_P(x)
=
\frac{1}{(2\pi\sigma^2)^{1/4}}
\exp\left(
-\frac{x^2}{4\sigma^2}+i\gamma x^2
\right),
\label{eq:chirped_gaussian}
\end{equation}
where $\gamma$ has dimensions of inverse length squared.

The quadratic phase does not change the position probability density, so the outcome distribution remains Eq.~\eqref{eq:gaussian_outcome_distribution}. However, the phase difference between the displaced and undisplaced amplitudes is
\begin{equation}
\begin{aligned}
\arg\psi_P(x-d)-\arg\psi_P(x)
&=\gamma(x-d)^2-\gamma x^2\\
&=\gamma d^2-2\gamma dx.
\end{aligned}
\label{eq:chirped_phase_difference}
\end{equation}
It follows that
\begin{equation}
\phi(x)=\phi_0+\gamma d^2-2\gamma dx
\pmod{2\pi}.
\label{eq:chirped_azimuthal_angle}
\end{equation}
The polar angle is still given by Eq.~\eqref{eq:gaussian_polar_angle}, while the azimuthal angle now varies with the measurement outcome. The conditional states therefore follow a nonmeridional trajectory on the Bloch sphere.

\begin{figure}[htbp]
\centering
\includegraphics[width=1.0\columnwidth]{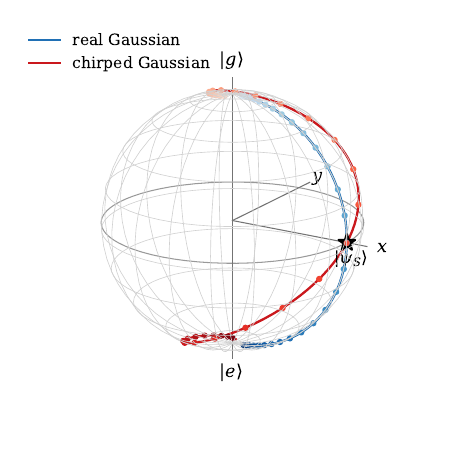}
\caption{
Conditional Bloch-sphere trajectories for a real Gaussian pointer
(red) and a chirped Gaussian pointer (blue); the star marks the
initial state $\ket{\psi_S}$. The real pointer keeps the conditional
states on the meridian $\phi=\phi_0$, whereas the chirped pointer
produces an outcome-dependent azimuthal angle and a nonmeridional
trajectory. The polar angle $\theta(x)$ is identical in both cases.
}
\label{fig:bloch_trajectories}
\end{figure}

Figure~\ref{fig:bloch_trajectories} illustrates the conditional trajectories on the Bloch sphere for both pointer preparations. The real Gaussian pointer confines the conditional states to the meridian $\phi=\phi_0$, while the chirped Gaussian pointer traces a nonmeridional path because the azimuthal angle $\phi(x)$ varies linearly with the measurement outcome $x$. The polar angle $\theta(x)$ is identical in the two cases, confirming that the amplitude profile of the pointer controls $\theta(x)$ whereas its phase profile controls $\phi(x)$.

The two angles are not independent controls: both are functions of the same measurement outcome $x$. The chirped pointer changes the trajectory traced by the conditional states, but it does not by itself allow the two angles to be independently prescribed for each shot.

More generally, a pointer wavefunction of the form
\begin{equation}
\psi_P(x)=A(x)e^{i\varphi(x)}
\label{eq:general_amplitude_phase}
\end{equation}
produces
\begin{equation}
\tan\frac{\theta(x)}{2}
=
\tan\frac{\theta_0}{2}
\frac{|A(x-d)|}{|A(x)|},
\label{eq:general_polar}
\end{equation}
and
\begin{equation}
\phi(x)=\phi_0+\varphi(x-d)-\varphi(x).
\label{eq:general_azimuthal}
\end{equation}
The amplitude and phase profiles therefore provide separate ways to shape the polar and azimuthal dependence of the conditional trajectory, although the two resulting coordinates remain linked through $x$.

\section{Outcome-dependent feedback and target-state preparation}

\subsection{Conditional state preparation}

The conditional state in Eq.~\eqref{eq:normalized_conditional_state} is pure for an ideal rank-one position measurement on the pointer, provided the initial qubit and pointer are pure and no additional decoherence is present. If the initial qubit state is known, an outcome-dependent unitary can map each conditional state to a prescribed target state.

Let the desired target be
\begin{equation}
\ket{\psi_{\mathrm{tar}}}
=
\cos\frac{\theta_{\mathrm{tar}}}{2}\ket{g}
+
e^{i\phi_{\mathrm{tar}}}
\sin\frac{\theta_{\mathrm{tar}}}{2}\ket{e}.
\label{eq:target_state}
\end{equation}
For each measurement outcome $x$, denote the normalized conditional state by $\ket{\psi_{S|x}}$. Choose normalized states orthogonal to the conditional and target states, denoted by $\ket{\psi_{S|x}^{\perp}}$ and $\ket{\psi_{\mathrm{tar}}^{\perp}}$, respectively. An outcome-dependent unitary can be constructed as
\begin{equation}
\begin{aligned}
\hat{V}_x
={}&
\ket{\psi_{\mathrm{tar}}}\bra{\psi_{S|x}}\\
&+
e^{i\chi_x}
\ket{\psi_{\mathrm{tar}}^\perp}
\bra{\psi_{S|x}^\perp},
\end{aligned}
\label{eq:feedback_unitary}
\end{equation}
where the phase $\chi_x$ can be chosen to fix the determinant convention. The operation satisfies
\begin{equation}
\hat{V}_x\ket{\psi_{S|x}}
=
\ket{\psi_{\mathrm{tar}}}.
\label{eq:feedback_action}
\end{equation}

In an ideal implementation, the measurement outcome is recorded and the corresponding unitary is applied. Since Eq.~\eqref{eq:feedback_unitary} depends on the known initial state through the conditional state, this protocol describes measurement-conditioned preparation of a target state. Its success as a practical control protocol depends on the ability to resolve the measurement outcome and implement the required feedback with sufficient accuracy.

\subsection{State preparation versus universal quantum gates}

It is essential to distinguish the above state-preparation procedure from a universal deterministic single-qubit gate. The conditional measurement operator is
\begin{equation}
\hat{M}_x
=
\begin{pmatrix}
\psi_P(x)&0\\
0&\psi_P(x-d)
\end{pmatrix}.
\end{equation}
For an arbitrary unknown input state, a deterministic unitary gate $\hat{U}_{\mathrm{tar}}$ would require a correction satisfying
\begin{equation}
\hat{V}_x\hat{M}_x
=
c_x\hat{U}_{\mathrm{tar}}
\label{eq:gate_condition}
\end{equation}
for every outcome with nonzero probability, where $c_x$ is a scalar amplitude. Because left multiplication by a unitary does not change the singular values of $\hat{M}_x$, this condition requires
\begin{equation}
|\psi_P(x)|=|\psi_P(x-d)|
\label{eq:equal_singular_values}
\end{equation}
for each relevant outcome, apart from the common scalar factor.

A localized Gaussian pointer does not satisfy this condition for all $x$ when $d\neq 0$. Therefore, the interaction and pointer-position measurement considered here do not, in general, implement an arbitrary deterministic unitary gate on an unknown qubit input through unitary feedforward alone.

This distinction matters for interpreting the protocol. Conditional state steering and feedback-based state preparation can be useful even when the measurement is nonunitary. Universal deterministic gate control, however, would require an additional protocol or a different measurement structure that preserves information about an arbitrary input state.

\section{Quantum Fisher information for parameter estimation}

\subsection{Joint qubit--pointer QFI}

The displacement parameter $d=\lambda\epsilon$ depends on the qubit energy separation $\epsilon$. We now evaluate the QFI of the joint pure state in Eq.~\eqref{eq:joint_state} with respect to $\epsilon$.

The parameter-dependent unitary is
\begin{equation}
\hat{U}_{\epsilon}
=
\exp\left(-\frac{i\epsilon\hat{G}}{\hbar}\right),
\qquad
\hat{G}=\lambda\ket{e}\bra{e}\otimes\hat{P}.
\label{eq:generator}
\end{equation}
For a pure state undergoing a unitary parameter encoding, the QFI is
\begin{equation}
\mathcal{F}_{Q}^{(S+P)}(\epsilon)
=
\frac{4}{\hbar^2}
\left(
\langle\hat{G}^2\rangle
-
\langle\hat{G}\rangle^2
\right).
\label{eq:pure_state_qfi}
\end{equation}
Define
\begin{equation}
s=\sin^2\frac{\theta_0}{2}.
\label{eq:excited_population}
\end{equation}
For the initial product state, the generator moments factorize as
\begin{equation}
\langle\hat{G}\rangle
=
\lambda s\langle\hat{P}\rangle,
\label{eq:generator_mean}
\end{equation}
and
\begin{equation}
\langle\hat{G}^2\rangle
=
\lambda^2s\langle\hat{P}^2\rangle.
\label{eq:generator_second_moment}
\end{equation}
Consequently,
\begin{equation}
\mathcal{F}_{Q}^{(S+P)}(\epsilon)
=
\frac{4\lambda^2}{\hbar^2}
\left[
s(\Delta P)^2+s(1-s)\langle\hat{P}\rangle^2
\right].
\label{eq:general_joint_qfi}
\end{equation}
For a pointer centered at zero momentum, $\langle\hat{P}\rangle=0$, this simplifies to
\begin{equation}
\mathcal{F}_{Q}^{(S+P)}(\epsilon)
=
\frac{4\lambda^2}{\hbar^2}
\sin^2\frac{\theta_0}{2}(\Delta P)^2.
\label{eq:joint_qfi}
\end{equation}

For the real Gaussian pointer in Eq.~\eqref{eq:gaussian_pointer}, using
$(\Delta P)^2=\hbar^2/(4\sigma^2)$ gives
\begin{equation}
\boxed{
\mathcal{F}_{Q}^{(S+P)}(\epsilon)
=
\frac{\lambda^2}{\sigma^2}
\sin^2\frac{\theta_0}{2}
}.
\label{eq:gaussian_joint_qfi}
\end{equation}
This is the QFI of the joint qubit--pointer state, not necessarily the QFI accessible from the pointer alone or the classical Fisher information of a position-only measurement.

For $\nu$ independent repetitions, the quantum Cram\'er--Rao bound for a locally unbiased estimator is
\begin{equation}
\Delta\epsilon
\geq
\frac{1}{\sqrt{\nu\mathcal{F}_{Q}^{(S+P)}(\epsilon)}}.
\label{eq:cramer_rao}
\end{equation}
The bound characterizes the information available in the joint state. Whether it can be attained using a particular measurement strategy depends on the measurement and estimation procedure.

\subsection{Effect of phase engineering and squeezing}

For the chirped Gaussian pointer in Eq.~\eqref{eq:chirped_gaussian}, the position variance remains $\sigma^2$, but the momentum variance is
\begin{equation}
(\Delta P)^2
=
\frac{\hbar^2}{4\sigma^2}
+
4\hbar^2\gamma^2\sigma^2.
\label{eq:chirped_momentum_variance}
\end{equation}
Substituting this expression into Eq.~\eqref{eq:joint_qfi} yields
\begin{equation}
\mathcal{F}_{Q}^{(S+P)}(\epsilon)
=
\lambda^2\sin^2\frac{\theta_0}{2}
\left(
\frac{1}{\sigma^2}
+
16\gamma^2\sigma^2
\right).
\label{eq:chirped_qfi}
\end{equation}
The additional term arises from the increased momentum variance associated with the quadratic phase. It should not be interpreted as a resource-free enhancement: phase engineering changes the pointer's momentum distribution.

Likewise, a narrower minimum-uncertainty Gaussian has a larger momentum variance and a larger joint-state QFI for this encoding. A fair comparison between pointer preparations should therefore specify the relevant resource constraint, such as momentum variance, mean energy, or an experimentally imposed bound on the available pointer states.

The joint-state QFI alone does not establish that a position measurement followed by feedback achieves the same precision. For a particular measurement with outcome distribution $p(x|\epsilon)$, the classical Fisher information is
\begin{equation}
\mathcal{F}_{\mathrm{cl}}(\epsilon)
=
\int_{-\infty}^{\infty}
\frac{1}{p(x|\epsilon)}
\left[
\frac{\partial p(x|\epsilon)}{\partial\epsilon}
\right]^2 dx,
\label{eq:classical_fisher}
\end{equation}
where the distribution is given by Eq.~\eqref{eq:outcome_probability} with $d=\lambda\epsilon$. This expression quantifies the information in the pointer-position record alone. If the conditional qubit is also measured, the full joint measurement statistics must be included in the Fisher-information calculation.
\begin{figure}[htbp]
\centering
\includegraphics[width=1.0\columnwidth]{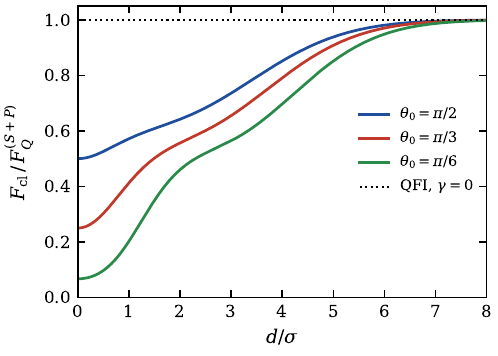}
\caption{
Ratio $\mathcal F_{\rm cl}(\epsilon)/\mathcal F_Q^{(S+P)}(\epsilon)$
of the position-measurement classical Fisher information to the
joint qubit--pointer QFI versus $d/\sigma$, for
$\theta_0=\pi/2$, $\pi/3$, and $\pi/6$, with a real Gaussian
pointer ($\gamma=0$). The ratio equals $\sin^2(\theta_0/2)$ at
$d=0$ and approaches unity as the displaced pointer states become
distinguishable. The dotted line marks $\mathcal F_{\rm cl}=\mathcal F_Q$.
}
\label{fig:cfi_qfi_ratio}
\end{figure}
Figure~\ref{fig:cfi_qfi_ratio} compares the classical Fisher
information of the position measurement with the joint-state QFI as a
function of $d/\sigma$ for several initial qubit states. At
$d/\sigma=0$ the two displaced pointer wavefunctions overlap
completely, the position record carries little information about
$\epsilon$, and the ratio reduces to $\sin^2(\theta_0/2)$. As
$d/\sigma$ increases, the two pointer states become distinguishable,
and the position measurement approaches the quantum Cramér--Rao bound.
The figure thus makes explicit that the joint-state QFI is an upper
bound that is approached only in the strong-displacement,
well-resolved regime, and that the fraction of the QFI accessible
from a position-only measurement depends on the initial qubit state.
\begin{figure}[htbp]
\centering
\includegraphics[width=0.9\columnwidth]{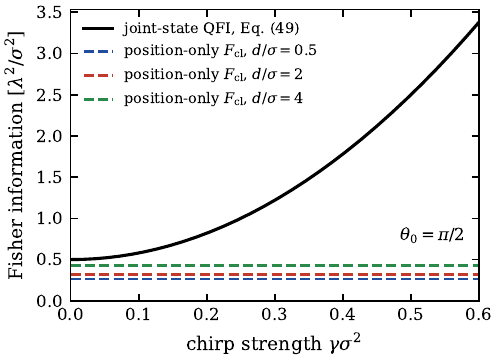}
\caption{
Joint-state QFI (black, Eq.~\eqref{eq:chirped_qfi}) and
position-only classical Fisher information (dashed) versus chirp
strength $\gamma\sigma^2$, for $\theta_0=\pi/2$ and
$d/\sigma=0.5,\,2,\,4$. The joint QFI grows quadratically with
$\gamma$, whereas the position-only CFI is independent of $\gamma$
because the quadratic phase leaves $|\psi_P(x)|^2$ unchanged.
}
\label{fig:qfi_chirp}
\end{figure}
Figure~\ref{fig:qfi_chirp} makes this limitation explicit for the
chirped pointer. The joint-state QFI (black) grows quadratically with
$\gamma$, consistent with the $16\gamma^2\sigma^2$ term in
Eq.~\eqref{eq:chirped_qfi}, while the position-only classical Fisher
information (dashed) is independent of $\gamma$ for every $d/\sigma$.
The reason is that the quadratic phase leaves $|\psi_P(x)|^2$
unchanged, so a position measurement cannot distinguish a chirped
pointer from an unchirped one. The additional QFI associated with
phase engineering is therefore not accessible through the pointer
position record alone and would require a different measurement, such
as one sensitive to the pointer momentum or to the joint
qubit--pointer state.

\section{Pointer-state superpositions from qubit postselection}

The same interaction can generate superpositions of displaced pointer states by measuring the qubit in a superposition basis. This procedure is distinct from projecting the pointer onto a superposition of its displaced states.

For example, consider a projective qubit measurement in the basis
\begin{equation}
\ket{\pm}
=
\frac{1}{\sqrt{2}}(\ket{g}\pm\ket{e}).
\label{eq:qubit_superposition_basis}
\end{equation}
Conditioned on outcome $+$ or $-$, the unnormalized pointer state is
\begin{equation}
\ket{\widetilde{\psi}_{P|\pm}}
=
\frac{1}{\sqrt{2}}
\left[
\cos\frac{\theta_0}{2}\ket{\psi_P}
\pm
e^{i\phi_0}\sin\frac{\theta_0}{2}
\ket{\psi_{P,d}}
\right].
\label{eq:postselected_pointer}
\end{equation}
After normalization, this is a coherent superposition of the undisplaced and displaced pointer states, provided both amplitudes are nonzero. The relative sign and phase depend on the qubit measurement outcome and the initial qubit phase.

The probability of the corresponding outcome is
\begin{equation}
p_{\pm}
=
\frac{1}{2}
\left\|
\cos\frac{\theta_0}{2}\ket{\psi_P}
\pm
e^{i\phi_0}\sin\frac{\theta_0}{2}
\ket{\psi_{P,d}}
\right\|^2.
\label{eq:postselection_probability}
\end{equation}
These probabilities include the overlap between the two pointer states. In particular, the displaced states are not generally orthogonal, so their superpositions must be normalized using their actual inner product.

Qubit postselection therefore provides a route to preparing pointer superpositions. The resulting state and its usefulness depend on the displacement, the initial qubit amplitudes, and the overlap between the pointer wavefunctions.

\section{Possible experimental implementation}

A possible implementation uses a qubit coupled to a motional mode, as in trapped-ion systems, or to a suitable bosonic mode in circuit quantum electrodynamics. State-dependent forces can generate conditional displacements of the mode, and measurements of the qubit or mode can provide the measurement record required for conditional preparation and feedback \cite{leibfried2003quantum,meekhof1996generation,wineland2003quantum,bruzewicz2019trapped}. In circuit QED, echoed conditional displacement sequences with dispersive qubit–cavity coupling provide an analogous resource \cite{eickbusch2022fast,ma2021quantum}.

For a trapped-ion mode, an interaction may be written schematically as
\begin{equation}
\hat{H}_{\mathrm{int}}
=
\hbar\chi\hat{\sigma}_z\otimes
\left(
\hat{a}e^{-i\phi_L}
+
\hat{a}^{\dagger}e^{i\phi_L}
\right),
\label{eq:ion_interaction}
\end{equation}
where \(\hat{a}\) and \(\hat{a}^{\dagger}\) are the mode annihilation and creation operators, \(\chi\) is the coupling rate, and \(\phi_L\) determines the coupled quadrature. By changing \(\phi_L\), the interaction can be made sensitive to different combinations of the mode's dimensionless quadratures. Recent work has demonstrated versatile spin-dependent-force generation and conditional displacement control across optical, metastable, and ground-state trapped-ion qubits \cite{bazavan2023synthesizing}.

However, Eq.~\eqref{eq:ion_interaction} contains a \(\hat{\sigma}_z\) coupling, whereas the desired interaction in Eq.~\eqref{eq:interaction_unitary} is proportional to the excited-state projector
\begin{equation}
\ket{e}\bra{e}
=
\frac{\hat{I}+\hat{\sigma}_z}{2}.
\label{eq:projector_decomposition}
\end{equation}
The desired coupling therefore contains both an identity-dependent pointer force and a qubit-dependent force. An experimental realization must demonstrate how these contributions are produced, calibrated, or otherwise accounted for. A spin-echo sequence cannot be assumed to remove the identity contribution while retaining the required qubit-dependent interaction without an explicit pulse-sequence derivation \cite{tyler2024laser}.

The practical protocol would require, at minimum, (i) preparation of the qubit and pointer states, (ii) a calibrated conditional displacement, (iii) a measurement that resolves the relevant pointer quadrature, and (iv) sufficiently fast and accurate outcome-dependent feedback \cite{manovitz2022trapped}. Decoherence, finite measurement resolution, imperfect displacement calibration, and feedback latency would all affect the achievable state fidelity. A quantitative experimental-feasibility claim therefore requires a platform-specific model and parameter estimates.

\section{Discussion}

The results establish a direct relationship between a pointer wavefunction and the trajectory of a conditionally measured qubit on the Bloch sphere. For a real Gaussian pointer, the relative phase between the two conditional amplitudes is constant, so the conditional state remains on a meridian. A phase-engineered pointer, such as a chirped Gaussian, introduces an outcome-dependent relative phase and changes the trajectory.

These trajectories describe conditional states rather than deterministic rotations generated by a unitary acting on an arbitrary unknown input. The measurement operator is generally nonunitary, and a unitary feedback operation cannot remove its unequal singular values for every outcome. The protocol is therefore most directly interpreted as a method for measurement-conditioned state steering and target-state preparation when the initial qubit state is known.

The QFI analysis also shows that parameter-estimation performance depends on the pointer's momentum variance. Narrowing the pointer or introducing a quadratic phase can increase the joint-state QFI, but these changes also modify the pointer's physical resources. A meaningful comparison with alternative protocols should specify the resource constraint and the measurement used to extract the parameter information.

The present formulation does not establish universal deterministic $SU(2)$ gate implementation or an advantage over existing measurement-based control methods. Establishing either result would require additional analysis. In particular, a universal-gate proposal would need a measurement and correction scheme that preserves arbitrary input-state information, while a metrological-advantage claim would require a resource-matched benchmark and a specified measurement strategy.

\section{Conclusion}

We have formulated a qubit--pointer interaction as a protocol for measurement-conditioned Bloch-sphere state steering. The conditional polar and azimuthal angles are determined by the amplitude and phase differences of the pointer wavefunction at positions separated by the conditional displacement. A real Gaussian pointer produces meridional trajectories, whereas a chirped Gaussian produces nonmeridional trajectories.

Outcome-dependent unitary feedback can prepare a prescribed target state from a known initial qubit state under ideal conditions. This state-preparation capability is distinct from universal deterministic quantum-gate implementation; the present conditional measurement operator does not by itself provide such a gate. We have also derived the joint-state QFI for estimating the displacement-generating parameter and shown explicitly how it depends on the pointer momentum variance.

These results provide a basis for studying conditional state preparation and measurement-assisted control with continuous-variable pointers. Further work should focus on finite-resolution measurements, realistic feedback errors, experimentally specified interactions, and comparisons with established control protocols under clearly defined resource constraints.

\subsection*{Declaration of AI Usage}

During the preparation of this manuscript, the author utilized an AI-based language model solely for the purposes of language refinement, grammar correction, and assistance with LaTeX formatting. The AI was not used to generate, modify, or interpret any scientific content, derivations, results, or conclusions presented in this work. All theoretical developments, analytical derivations, numerical calculations, and scientific interpretations were carried out independently by the author. The author takes full responsibility for the accuracy and integrity of the content presented in this manuscript.

\end{document}